# Graviton Dominated Eras of Universe's Evolution, Inflation and Dark Energy

Leonid Marochnik

Physics Department, East-West Space Science Center, University of Maryland, College Park, MD 20742

**Abstract**

The empty space (with no matter fields) is not really empty because of natural metric fluctuations, quantum (gravitons) and classical (gravitational waves). We show that gravitons as well as classical gravitational waves of super-horizon wavelengths are able to form the de Sitter state of the empty homogeneous isotropic Universe. This state is the exact solution to the self-consistent equations of finite one-loop quantum gravity for gravitons in the empty FLRW space. It also is the exact solution to the self-consistent equations of back reaction for classical gravitational waves in the same space. Technically, to get this de Sitter solution in both quantum and classical cases, it is necessary to make the transition to imaginary time and then come back to real time which is possible because this de Sitter state is invariant with respect to Wick rotation. Such a procedure means that the time was used as a complex variable, and this fact has a deep but still not understood meaning. De Sitter accelerated expansion of the empty Universe naturally explains the origin of dark energy and inflation because the Universe is empty at the start (inflation) and by the end (dark energy) of its evolution. This theory is consistent with the existing observational data. The CMB anisotropy of the order of 10^-5 is produced by fluctuations in the number of gravitons. The existence of a threshold and unique coincidence of topologically impenetrable barriers for tunneling takes place for the matter-dominated epoch and De Sitter State only. These facts provide a solution to the coincidence problem. The theoretical prediction that the equation-of-state parameter should be $w > -1$ for inflation and $w < -1$ for dark energy is consistent with observational data. To provide the reader with a complete picture, this paper gather together new and some published results of the graviton theory of the origin of inflation and dark energy.

## 1. Introduction

Inflation and dark energy are two unsolved puzzles of modern cosmology. The idea of the necessity of inflation does not yet have direct reliable observational confirmation but it seems very attractive due to the ability to solve three known major cosmological paradoxes which are flatness, horizon and monopoles [1]. Nevertheless, it is useful to remind here of Weinberg's remark [2] that "so far, the details of inflation are unknown and the whole idea of inflation remains a speculation, though one that is increasingly plausible". On the other hand, the existence of dark energy is an established observational fact [3, 4]. In case of inflation, it is almost generally accepted that the acceleration of expansion is due to a highly hypothetical scalar field. In case of dark energy, the cosmological constant is considered as the most popular candidate to provide the acceleration. The equation of state of dark energy is unknown, and it is usual to use an equation-of-state parameter $w = p/\varepsilon$ as a measure of proximity to $\Lambda$. At the present time, $w \approx -1$ is consistent with observations [5, 6]. However, the observational constraint on the equation-of-state parameter of dark energy $w \approx -1$ tells us only that the equation of state of dark energy is close to $p = -\varepsilon$, and this does not mean that it is necessarily due to $\Lambda \neq 0$. Such an equation of state can exist due to other reason (section 4). In addition, there are well-known unsolved problems if one considers $\Lambda$ as the cause of the dark energy effect. The first one is a so-called "old cosmological constant problem": Why is the $\Lambda$-value measured from observations is of the order of $10^{-122}$ vacuum energy density? The second one is a "coincidence problem": Why is the acceleration happening during the contemporary epoch of matter domination? As to a hypothetical scalar field as a cause of inflation, we show that it is not necessary to make any hypotheses (including the scalar field) to generate inflation. Both inflation and dark energy are tied by the fact that in both cases we are dealing with an exponentially accelerated expansion of the Universe (for similarities between primordial acceleration and present dark energy see, e.g. [7]). Meantime, there is one more important feature uniting them, which is the emptiness of the Universe. By the end of its evolution, the Universe is going to become empty because the energy density of non-relativistic matter $\varepsilon_m \sim a(t)^{-3} \to 0$ as $t \to \infty$ where $a$ is its scale factor. Currently, it is almost empty being filled with such matter of the order of 25-30% [6], so that the effects associated with the emptiness of space must already be very noticeable, and they are (dark energy). It is usual to call space without matter fields empty. Hereinafter, the term "empty space" is used in this sense. At the start of cosmological evolution before matter was born, the Universe is also assumed to be empty. Both observational data on dark energy tell us that the almost empty Universe is expanding with acceleration and the necessity of inflation ("increasingly plausible speculations") tells us that the empty new-born Universe is also expanding with acceleration. So, both phenomena are somehow associated with the emptiness of the Universe which is not a coincidence. The empty space is not really empty because of natural metric fluctuations, quantum (gravitons) and classical (gravitational waves). We show that the de Sitter accelerated expansion of the empty Universe is its natural state, and there is no need for additional hypotheses to explain the origin of inflation and dark energy. It seems natural to propose to add to the generally accepted history of the evolution of the Universe two additional stages which are "graviton dominated Universe" at the beginning and "gravitational wave dominated Universe" at the end of its evolution.

As is known, the idea of a possible key role of de Sitter solution for the early stages of evolution of the Universe goes back to the "founding fathers" of modern cosmology, de Sitter, Eddington and Lemaitre. Gliner [8] was the first to show that the equation of state $p = -\varepsilon$, generating de Sitter expansion of the Fiedmanian Universe is the equation of state of anti-gravitating physical vacuum. On the basis of this idea, Zeldovich [9] showed that the cosmological constant producing the de Sitter expansion of the Friedmanian Universe is of the vacuum nature. In the pioneering work of Starobinsky [10], it was shown that the quantum corrections to Einstein's equations that come from conformal anomalies can lead to the nonsingular de Sitter solution instead of the standard Big Bang singularity. The possibility of de Sitter expansion at the beginning of cosmological evolution was also found in [11, 12]. In the present work, we show that the de Sitter state is formed by back reaction of quantum metric fluctuations (gravitons) as well as classical gravitational waves (CGW) in the empty homogeneous isotropic Universe in imaginary time. We show that this de Sitter state is invariant with respect to Wick rotation $t = i\tau$. The latter suggests that it takes place in real time too. By its very nature, the de Sitter expansion of the early empty Universe formed by gravitons is a quantum coherent condensate [15]. The de Sitter expansion of the late emptying Universe can be formed by CGW as well as quantum gravitons, both of super-horizon wavelengths. Despite the intuitive feeling that quantum effects may not be significant in the modern Universe, this fact holds because we deal here with a macroscopic quantum effect similar to superconductivity and superfluidity [15]. To the idea of "founding fathers'" on the key role of de Sitter expansion for early stages of evolution, we add the idea of the key role of the de Sitter expansion not only for the early stages of evolution (inflation) but also for the late stages of evolution of the Universe (dark energy). This paper represents an attempt to gather together new and published results with emphasis on their cosmological applications (the effects of inflation and dark energy). As such it leads to a certain overlap with our previously published works [13-18]. To make it easier to read this paper, we found it necessary to go beyond references to our published work and included in the paper some already published results along with yet unpublished results.

The outline of the paper is the following. In section 2, we show that the virtual gravitons in the empty FLRW space described by equations of one-loop finite quantum gravity form the de Sitter state. To get this de Sitter solution for the empty FLRW space, we used the direct and then reverse Wick rotation. In section 3, we confirmed this result (formation of de Sitter state) by obtaining it by the independent method of BBGKY chain. We show that in this independent approach it is also necessary to make the Wick rotation and then come back to real time to get the de Sitter solution for the empty FLRW space. In section 4, we show that the back reaction of classical gravitational waves on the WLRW background also form the de Sitter state but again it is necessary to make the Wick rotation and then come back to real time to get this result. In section 5, we consider the gravitons and classical gravitational waves forming the self-consistent de Sitter state in terms of tunneling . In section 6, we show that the de Sitter solution obtained this way is consistent with all existing observational data on inflation and dark energy. The conclusion is given in section 7.

**2. Self-consistent theory of gravitons in macroscopic FLRW space; the de Sitter solution**

The exact equations of self-consistent theory of gravitons in the Heisenberg representation with the ghost sector automatically providing one-loop finiteness off the mass shell are obtained in our work [16]. In [16], it is shown that *the Heisenberg representation of quantum gravity* (*as well as the Heisenberg representation of quantum Yang-Mills theory* [19]) *exists only in the Hamilton gauge. The ghost sector corresponding to this gauge is represented by the complex scalar field with minimal coupling to gravity*. These equations are the only mathematically consistent equations in the available literature [16][1]. In [16], the exact equations in the Heisenberg representation in the Hamilton gauge together with postulates of canonical quantization claim the status of the theory formulated only on the basis of first principles of quantum gravity. The approximate theory of macroscopic system of gravitons in the macroscopic space-time with the self–consistent geometry can be obtained from these equations. In this theory, the interaction of gravitons with the classical gravitational field is taken into account exactly, and the perturbation theory over the amplitude of quantum fluctuations is only used when describing the graviton–graviton interaction. Referring the reader to the works [16], [13] and [15] for details of the transition from the exact equations in the Heisenberg representation in the Hamilton gauge to the one-loop approximation, we present at once the final equations of one-loop approximation in the FLRW metric. The equations of one-loop quantum theory of gravitons in the empty flat FLRW space-time are the following. The background metric is described by regular Einstein equations

---

[1] Unfortunately, the vast majority of papers on quantum theory of gravitons published in 1977-2008 were incorrect (see [16] and references therein). All these works lose any meaning after renormalization of gravitational Lagrangian by quadratic counterterms off the mass shell. This procedure modifies the original definition of the graviton. This fact makes such works mathematically inconsistent and physically meaningless. In the finite one-loop quantum theory of gravitons presented in [13, 16], the incorrect counterterms simply do not arise. In other words, the renormalization procedure (used in all papers known to us) is inapplicable to the quantum theory of gravitons in the one-loop approximation. Also, the linear parameterization used in all these works (with no exception) also inapplicable to the theory of gravitons of long wavelengths in the one-loop approximation. This fact was documented in work [28]. As it was shown in [13, 16], the exponential parameterization corresponding to normal coordinates in the functional space solves the problem (this procedure is used in this paper). This fact applies to both quantum gravitons and classical gravitational waves [13, section II.D].

$$R_i^k - \frac{1}{2}\delta_i^k R = \kappa \left( <\Psi_g | \hat{T}_{i(grav)}^k | \Psi_g > + <\Psi_{gh} | \hat{T}_{i(ghost)}^k | \Psi_{gh} > \right) \quad (1)$$

The energy-momentum tensor (EMT) of gravitons $\hat{T}_{i(grav)}^k$ and ghosts $\hat{T}_{i(ghost)}^k$ were obtained by solving operator equations of motion and averaging over a quantum ensemble $|\Psi>$. The explicit form of $\hat{T}_{i(grav)}^k$ and $\hat{T}_{i(ghost)}^k$ can be found in [16, 13]. In the empty flat isotropic and homogeneous Universe, the Einstein equations (1) read [15]

$$3H^2 = \kappa \varepsilon_g \quad (2)$$

$$2\dot{H} + 3H^2 = -\kappa p_g \quad (3)$$

$$\varepsilon_g = \frac{1}{8}\sum_{\mathbf{k}\sigma}<\Psi_g | \dot{\hat{\psi}}_{\mathbf{k}\sigma}^+ \dot{\hat{\psi}}_{\mathbf{k}\sigma} + \frac{k^2}{a^2}\hat{\psi}_{\mathbf{k}\sigma}^+ \hat{\psi}_{\mathbf{k}\sigma} | \Psi_g > - \frac{1}{4}\sum_{\mathbf{k}}<\Psi_{gh} | \dot{\hat{\theta}}_{\mathbf{k}}^+ \dot{\hat{\theta}}_{\mathbf{k}} + \frac{k^2}{a^2}\hat{\theta}_{\mathbf{k}}^+ \hat{\theta}_{\mathbf{k}} | \Psi_{gh} >,$$

$$p_g = \frac{1}{8}\sum_{\mathbf{k}\sigma}<\Psi_g | \dot{\hat{\psi}}_{\mathbf{k}\sigma}^+ \dot{\hat{\psi}}_{\mathbf{k}\sigma} - \frac{k^2}{3a^2}\hat{\psi}_{\mathbf{k}\sigma}^+ \hat{\psi}_{\mathbf{k}\sigma} | \Psi_g > - \frac{1}{4}\sum_{\mathbf{k}}<\Psi_{gh} | \dot{\hat{\theta}}_{\mathbf{k}}^+ \dot{\hat{\theta}}_{\mathbf{k}} - \frac{k^2}{3a^2}r\hat{\theta}_{\mathbf{k}}^+ \hat{\theta}_{\mathbf{k}} | \Psi_{gh} > \quad (4)$$

Where $\Psi_g$ and $\Psi_{gh}$ are quantum state vectors of gravitons and ghosts, respectively; $H = \dot{a}/a$; $a(t)$ is the scale factor of a flat FLRW model; $\kappa = 8\pi G$; speed of light $c = 1$; dots are time derivatives; $\sigma$ is the polarization index and superscript "+" denotes complex conjugation. Heisenberg's operator equations for Fourier components of the transverse 3–tensor graviton field $\hat{\psi}_{\mathbf{k}\sigma}$ and Grassman ghost field $\hat{\theta}_{\mathbf{k}}$ are:

$$\ddot{\hat{\psi}}_{\mathbf{k}\sigma} + 3H\dot{\hat{\psi}}_{\mathbf{k}\sigma} + \frac{k^2}{a^2}\hat{\psi}_{\mathbf{k}\sigma} = 0 \quad (5)$$

$$\ddot{\hat{\theta}}_{\mathbf{k}} + 3H\dot{\hat{\theta}}_{\mathbf{k}} + \frac{k^2}{a^2}\hat{\theta}_{\mathbf{k}} = 0 \quad (6)$$

Canonical commutation relations for gravitons and anti-commutation relations for ghosts read

$$\frac{a^3}{4}\left[\dot{\hat{\psi}}_{\mathbf{k}\sigma}^+, \hat{\psi}_{\mathbf{k}\sigma}\right]_- = -i\hbar \delta_{\mathbf{k}\mathbf{k}'}\delta_{\sigma\sigma'},$$

$$\frac{a^3}{8}\left[\dot{\hat{\theta}}_{\mathbf{k}}^+, \hat{\theta}_{\mathbf{k}}\right]_+ = -\frac{a^3}{8}\left[\dot{\hat{\theta}}, \hat{\theta}_{\mathbf{k}}^+\right]_+ = -i\hbar \delta_{\mathbf{k}\mathbf{k}'}. \quad (7)$$

Equations (5) and (6) and quantization rules (7) have been obtained by Faddeev & Popov's path integral from the class of synchronic gauges that automatically provide one–loop finiteness of observables (see [13] section III.D). One–loop effects of vacuum polarization and particle creation by the background field are contained in equations (5-6) for gravitons and ghosts. These equations are linear in quantum fields but their coefficients depend on the non–stationary background metric. Correspondingly, in the background equation (2-3) we keep the average values of bilinear forms of quantum fields only. In this model, quantum particles interact through a common self–consistent field only. We took also into account the following definitions

$$<\Psi|\hat{T}_0^0|\Psi> = \varepsilon_g, \quad <\Psi|\hat{T}_\alpha^\beta|\Psi> = \frac{\delta_\alpha^\beta}{3}<\Psi|\hat{T}_\gamma^\gamma|\Psi> = -\delta_\alpha^\beta p_g \tag{8}$$

The mathematically rigorous method of separating the background and the gravitons, which ensures the existence of the EMT, is based on averaging over graviton polarizations: $<\psi_\alpha^\beta> \equiv 0$ if all polarizations are equivalent in the quantum ensemble[2]. Equations (2-6) form the self-consistent system of equations of one-loop quantum gravity for gravitons and FLRW background. In equations (2-4), it is convenient to use the conformal $\eta = \int dt/a$ and to go from summation to integration by the transformation $\sum_\mathbf{k} ... \to \int d^3k/(2\pi)^3 ... = \int_0^\infty k^2 dk/2\pi^2 .....$ From (2-4) follow the Friedmanian equations for the energy density and pressure [13, 17]

$$3\frac{a'^2}{a^4} = \kappa\varepsilon_g = \frac{1}{16\pi^2}\int_0^\infty \frac{k^2}{a^2}dk(\sum_\sigma <\Psi_g|\hat{\psi}'^+_{\mathbf{k}\sigma}\hat{\psi}'_{\mathbf{k}\sigma} + k^2\hat{\psi}^+_{\mathbf{k}\sigma}\hat{\psi}_{\mathbf{k}\sigma}|\Psi_g> \\ -2<\Psi_{gh}|\bar{\theta}'_\mathbf{k}\theta'_\mathbf{k} + k^2\bar{\theta}_\mathbf{k}\theta_\mathbf{k}|\Psi_{gh}>) \tag{9}$$

$$2\frac{a''}{a^3} - \frac{a'^2}{a^4} = -\kappa p_g = -\frac{1}{16\pi^2}\int_0^\infty \frac{k^2}{a^2}dk(\sum_\sigma <\Psi_g|\hat{\psi}'^+_{\mathbf{k}\sigma}\hat{\psi}'_{\mathbf{k}\sigma} \\ -\frac{k^2}{3}\hat{\psi}^+_{\mathbf{k}\sigma}\hat{\psi}_{\mathbf{k}\sigma}|\Psi_g> -2<\Psi_{gh}|\bar{\theta}'_\mathbf{k}\theta'_\mathbf{k} - \frac{k^2}{3}\bar{\theta}_\mathbf{k}\theta_\mathbf{k}|\Psi_{gh}>) \tag{10}$$

Equations (5-6) in conformal time read

$$\hat{\phi}''_{\vec{k},\sigma} + (k^2 - \frac{a''}{a})\hat{\phi}_{\vec{k},\sigma} = 0 \quad \hat{\psi}_{\mathbf{k}\sigma} = \frac{1}{a}\hat{\phi}_{\mathbf{k}\sigma} \tag{11}$$

$$\hat{\vartheta}''_\mathbf{k} + (k^2 - \frac{a''}{a})\hat{\vartheta}_\mathbf{k} = 0 \quad \hat{\theta}_\mathbf{k} = \frac{1}{a}\hat{\vartheta}_\mathbf{k} \tag{12}$$

The self-consistent set of equations of one-loop quantum gravity which are finite off the mass shell is formed by (9-12). One can see from (9-12) that in the mathematical formalism of the theory, the ghosts play a role of a second physical subsystem, whose average contributions to the macroscopic Einstein equations appear on an equal footing with the average contribution of gravitons. At the first glance, it may seem that the status of the ghosts as the second subsystem is in contradiction with the well–known fact that the Faddeev–Popov ghosts are not physical particles. However the paradox is in the fact that we have no contradiction with the standard concepts of quantum theory of gauge fields but rather full agreement with these. The Faddeev–Popov ghosts are indeed not physical particles in a quantum–field sense, that is, they are not particles that are in the asymptotic states whose energy and momentum are connected by a definite relation. Such ghosts are nowhere to be found on the pages of our work. We discuss only virtual gravitons and virtual ghosts that exist in the area of interaction. As to virtual ghosts, they cannot be eliminated in principle due to lack of ghost–free gauges in quantum gravity. In the strict mathematical sense, the non–stationary Universe as a whole is a region of interaction, and, formally speaking, there are no real gravitons and ghosts in it. To finally clarify the problem of the presence of ghosts in the equations of the theory, we quote a footnote #2 from our work in [13][3] (see also [16]).

---

[2] There are also other ways of averaging and in particular, the averaging over 3-space. For the first time, this method was used in [20] and then in [21]

The de Sitter solution in terms of conformal time reads
$$a = -(H\eta)^{-1} \qquad (13)$$
The solution of (11-12) over the background (2.13) reads

$$\hat{\psi}_{\mathbf{k}\sigma} = \frac{1}{a_S}\sqrt{\frac{2\kappa\hbar}{k}}\left[\hat{c}_{\mathbf{k}\sigma}f(x) + \hat{c}^+_{-\mathbf{k}-\sigma}f^+(x)\right],$$

$$\hat{\theta}_{\mathbf{k}} = \frac{1}{a_S}\sqrt{\frac{2\kappa\hbar}{k}}\left[\hat{\alpha}_{\mathbf{k}}f(x) + \hat{\bar{\beta}}_{-\mathbf{k}}f^+(x)\right], \qquad (14)$$

$$f(x) = (1 - \frac{i}{x})e^{-ix} \qquad x = k\eta \qquad (15)$$

Substitution of (13), (14) and (15) into (9-12) leads to divergent integrals in RHS of (9-10) [17]. The problem can be solved by Wick rotation $t = i\tau$ and $\eta = i\varsigma$. In imaginary conformal time $\varsigma$, equations (9), (11) and (13) read [17]

---

[3] "We emphasize that the equal participation of virtual gravitons and ghosts in the formation of macroscopic observables in the non–stationary Universe does not contradict the generally accepted concepts of the quantum theory of gauge fields. On the contrary it follows directly from the mathematical structure of this theory. In order to clear up this issue once and for all, recall some details of the theory of S− matrix. In constructing this theory, all space–time is divided into regions of asymptotic states and the region of effective interaction. Note that this decomposition is carried out by means of, generally speaking, an artificial procedure of turning on and off the interaction adiabatically. (For obvious reasons, the problem of self–consistent description of gravitons and ghosts in the non-stationary Universe with $\lambda H \approx 1$ by means of an analogue of such procedure cannot be considered a priori.) Then, after splitting the space–time into two regions, it is assumed that the asymptotic states are ghost–free. In the most elegant way, this selection rule is implemented in the BRST formalism, which shows that the BRST invariant states turn out to be gauge–invariant automatically. The virtual ghosts, however, remain in the area of interaction, and this points to the fact that virtual gravitons and ghosts are parts of the Feynman diagrams on an equal footing. In the self–consistent theory of gravitons in the non–stationary Universe, virtual ghosts of equal weight as the gravitons, appear at the same place where they appear in the theory of S−matrix, i.e. at the same place as they were introduced by Feynman, i.e. in the region of interaction. Of course, the fact that in the real non–stationary Universe, both the observer and virtual particles with $\lambda H \approx 1$ are in the area of interaction, is highly nontrivial. It is quite possible that this property of the real world is manifested in the effect of dark energy. An active and irremovable participation of virtual ghosts in the formation of macroscopic properties of the real Universe poses the question of their physical nature. Today, we can only say with certainty that the mathematical inevitability of ghosts provides the one–loop finiteness off the mass shell, i.e. the mathematical consistency of one-loop quantum gravity without fields of matter".

$$-3\frac{a'^2}{a^4} = \frac{1}{16\pi^2}\int_0^\infty \frac{k^2}{a^2}dk(\sum_\sigma <\Psi_g|-\hat{\psi}'^+_{k\sigma}\hat{\psi}'_{k\sigma}+k^2\hat{\psi}^+_{k\sigma}\hat{\psi}_{k\sigma}|\Psi_g>$$
$$-2<\Psi_{gh}|-\bar{\theta}'_k\theta'_k+k^2\bar{\theta}_k\theta_k|\Psi_{gh}>) \tag{16}$$

$$\hat{\phi}''_{k,\sigma} - (k^2 + \frac{a''}{a})\hat{\phi}_{k,\sigma} = 0 \qquad \hat{\psi}_{k\sigma} = \frac{1}{a}\hat{\phi}_{k\sigma} \tag{17}$$

$$\hat{\vartheta}''_k - (k^2 + \frac{a''}{a})\hat{\vartheta}_k = 0 \qquad \hat{\theta}_k = \frac{1}{a}\hat{\vartheta}_k \tag{18}$$

$$a = -(H_\tau\varsigma)^{-1} \tag{19}$$

$$H_\tau = \frac{1}{a}\frac{da}{d\tau} = iH \tag{20}$$

Where

$$\hat{\psi}_{k\sigma} = \frac{1}{a}\hat{\phi}_{k\sigma}, \quad \hat{\theta}_k = \frac{1}{a}\hat{\vartheta}_k.$$

Primes in this section denote derivatives over imaginary conformal time $\varsigma$. The solutions to (17-18) over the de Sitter background (19) read

$$\hat{\psi}_{k\sigma} = \frac{1}{a}\sqrt{\frac{2\kappa\hbar}{k}}\left(\hat{Q}_{k\sigma}g_k + \hat{P}_{k\sigma}h_k\right) \qquad \hat{\theta}_k = \frac{1}{a}\sqrt{\frac{2\kappa\hbar}{k}}\left(\hat{q}_k g_k + \hat{p}_k h_k\right) \tag{21}$$

Where

$$g(\xi) = \left(1+\frac{1}{\xi}\right)e^{-\xi}, \qquad h(\xi) = \left(1-\frac{1}{\xi}\right)e^{\xi} \qquad \xi = k\varsigma \tag{22}$$

For the imaginary time formalism and operator constants of integration $\hat{Q}_{k\sigma}, \hat{P}_{k\sigma}, \hat{q}_k, \hat{p}_k$ see [13, section VII][4]. The requirement of finiteness eliminates the $h-$ solution[5]. After such a transition to imaginary time, the substitution of (19), (21) and (22) into (16-18) leads to the following equation for $H_\tau$ [17] (the graviton spectrum here is supposed to be flat, i.e. $N_k = N = const$)

$$H_\tau^2 = -\frac{\kappa \hbar H_\tau^4 N}{8\pi^2} \qquad (23)$$

---

[4] For imaginary time formalism and operator constants of integration $\hat{Q}_{k\sigma}, \hat{P}_{k\sigma}, \hat{q}_k, \hat{p}_k$ see [13, section VII]

Where $N_k = \sum_\sigma <\Psi_g | \hat{Q}^+_{k\sigma} \hat{Q}_{k\sigma} | \Psi_g> - 2<\Psi_{gh} | \hat{q}^+_k \hat{q}_k | \Psi_{gh}>$ and $N$ is the number of gravitons in the space (the details of calculations can be found in [13, 17]). The eq. (2) in imaginary time reads

$$\kappa \varepsilon_\tau = -3H_\tau^2 \qquad (24)$$

Substituting (23) into (24) we get the equation of state $p_\tau = p_\tau(\varepsilon)$ in imaginary time that reads

$$-p_\tau = \varepsilon_\tau = \frac{3\hbar N}{8\pi^2} H_\tau^4 \qquad (25)$$

Using (20), one can get the solution to (23) in real time that reads

$$H^2 = \frac{8\pi^2}{\kappa \hbar N} \qquad (26)$$

The de Sitter state is invariant with respect to Wick rotation. This can be seen from the following

$$a = \exp(H_\tau \cdot \tau) = \exp(iH \cdot -it) = \exp(H \cdot t) \qquad (27)$$

The invariance (27) suggests that the de Sitter solution obtained in imaginary time is valid also in real time. To get the equation of state in real time, one needs to change $H_\tau \to H$ in the equation of state (25). After this substitution, one can see that the equation of state (25) is also invariant with respect to Wick rotation. It follows from the fact that

$$H^4 = (-iH_\tau)^4 = H_\tau^4 \qquad (28)$$

So we get the same equation of state as (25) $p_g = -\varepsilon_g$ that reads

---

[5] As is shown in [13] (section VII.B), the requirement $\hat{P}_{k\sigma} = \hat{p}_k = 0$ needed to eliminate $h$ solution, leads to the fact that the *graviton-ghost system forms a quantum coherent instanton condensate.* The effect of dark energy is observed in the contemporary Universe which is far from the Planck time. Therefore, quantum origin of it seems counterintuitive. In fact, this is a macroscopic quantum effect similar to superconductivity and superfluidity [13]. Its origin is related to the formation of quantum coherent condensate. Due to the one–loop finiteness of self–consistent theory of gravitons, all observables are formed by the difference between graviton and ghost contributions. This fact can be seen from the definition of $N_k$. The same final differences of contributions may correspond to the totally different graviton and ghost contributions themselves. All quantum states are degenerate with respect to mutually consistent transformations of gravitons and ghost's occupation numbers, but providing unchanged values of observable quantities. *This is a direct consequence of the internal mathematical structure of the self–consistent theory of gravitons, satisfying the one–loop finiteness condition.* The tunneling that unites degenerate quantum states into a single quantum state produces a quantum coherent instanton condensate in imaginary time. We refer the reader to [13, section VII] for details. In a general form, hypotheses on the possibility of graviton condensate formation in the Universe were proposed in [22] and [23] (see [13] for more details).

$$-p_g = \varepsilon_g = \frac{3\hbar N}{8\pi^2} H^4 \qquad (29)$$

To conclude this section, note that the Wick rotation applied to the self-consistent set of equations of one-loop finite quantum gravity for the empty space leads to the de Sitter solution for the real time empty Universe. By its nature, this de Sitter solution is a quantum coherent condensate that produces a macroscopic quantum effect of cosmological expansion that can be responsible for inflation and dark energy effects.

### 3. De Sitter state of empty space as exact solution to BBGKY chain

In view of the fundamental importance of the existence of the de Sitter solution for empty space, we obtain this solution by yet another independent method. In this section, we obtain the same de Sitter solution for the empty FLRW space by means of Bogolyubov-Born-Green-Kirkwood-Yvon hierarchy (BBGKY chain). For the first time, it was obtained in work [14]. To build the BBGKY chain, one needs to introduce the graviton spectral function $W_{\mathbf{k}}$ and its moments $W_n$

$$W_{\mathbf{k}} = \sum_\sigma <\Psi_g | \hat{\psi}^+_{\mathbf{k}\sigma} \hat{\psi}_{\mathbf{k}\sigma} | \Psi_g> - 2 <\Psi_{gh} | r\theta_{\mathbf{k}}\theta_{\mathbf{k}} | \Psi_{gh}>$$

$$W_n = \sum_{\mathbf{k}} \frac{k^{2n}}{a^{2n}} \left( \sum_\sigma <\Psi_g | \hat{\psi}^+_{\mathbf{k}\sigma} \hat{\psi}_{\mathbf{k}\sigma} | \Psi_g> - 2 <\Psi_{gh} | \bar{\theta}_{\mathbf{k}}\theta_{\mathbf{k}} | \Psi_{gh}> \right) \qquad n=0,1,2,...,\infty \quad (30)$$

The derivation of BBGKY chain can be found in [13, section V]. It reads

$$\dot{D} + 6HD + 4\dot{W}_1 + 16HW_1 = 0 \qquad (31)$$

$$\dddot{W}_n + 3(2n+3)H\ddot{W}_n + 3\left[ \left(4n^2+12n+6\right)H^2 + (2n+1)\dot{H} \right]\dot{W}_n +$$
$$+ 2n\left[ 2\left(2n^2+9n+9\right)H^3 + 6(n+2)H\dot{H} + \ddot{H} \right]W_n + 4\dot{W}_{n+1} + 8(n+2)HW_{n+1} = 0 \qquad n=1,...,\infty \quad (32)$$

$$D = \ddot{W}_0 + 3H\dot{W}_0 \qquad (33)$$

Equations (31) and (32) form the BBGKY chain. Each equation of this chain connects the neighboring moments. Equations (31) and (32) have to be solved jointly with the Einstein equations (2-3). In terms of $D$ and $W_1$ the energy density and pressure of gravitons are

$$3H^2 = \kappa \varepsilon_g \equiv \frac{1}{16}D + \frac{1}{4}W_1, \qquad 2\dot{H} + 3H^2 = -\kappa p_g \equiv \frac{1}{16}D + \frac{1}{12}W_1, \qquad (34)$$

Instead of the original self-consistent system of (2-4), (6) and (7) we get now the new self-consistent system of equations consisting of the Einstein equations (34) and BBGKY chain (31), (32). The energy-momentum tensor (34) can be reduced to the form found in [21] by identity transformations. As was shown in [14], the de Sitter solution is one of exact solutions to the equations of BBGKY chain for the empty FLRW space. One can check by simple substitution that it reads

$$H^2 = \frac{1}{36}W_1, \quad \varepsilon_g = -p_g = \frac{1}{12\kappa}W_1, \quad D = -\frac{8}{3}W_1, \quad a = a_0 e^{Ht} \tag{35}$$

$$W_{n+1} = -\frac{n(2n+3)(n+3)}{2(n+2)}H^2 W_n, \quad n \geq 1 \tag{36}$$

In terms of "microscopic" calculations (section 4) $D$ and $W_n$ read [13]

$$D = -\frac{12\hbar N}{\pi^2}H^4, \quad W_n = \frac{(-1)^{n+1}}{2^{2n}}(2n-1)!(2n+1)(n+2) \times \frac{2\hbar N}{\pi^2}H^{2n+2}, \quad n \geq 1, \tag{37}$$

$$H = const; \quad W_n = const; \quad D = const.$$

The zero moment $W_0$, which has an infrared logarithmic singularity, is not contained in the expressions for the macroscopic observables, and for that reason, is not calculated. In the equation for $W_0$, the functions are differentiated in the integrand and the derivatives are combined in accordance with the definition (33). At the last step the integrals that are calculated, already possess no singularities.

From (36-37) it follows that the solution has essentially an instanton nature. It can be seen from the fact that the signs of the moments $W_{n+1}/W_n < 0$ alternate. This alternation means that even moments are negative. Transition to imaginary time removes alternation of the moments and leads to the following changes

$$t = i\tau; \quad H = -iH_\tau; \quad H_\tau = \frac{1}{a}\frac{da}{d\tau}; \quad D \rightarrow -D^\tau; \quad D^\tau = \frac{d^2}{d\tau^2}W_0 + 3H_\tau\frac{d}{d\tau}W_0$$

$$-H_\tau^2 = \frac{1}{36}W_1; \quad D_\tau = \frac{8}{3}W_1 \tag{38}$$

$$W_{n+1} = \frac{n(2n+3)(n+3)}{2(n+2)}H_\tau^2 W_n, \quad n \geq 1;$$

$$a = a_0 e^{H_\tau \tau} \qquad (39)$$

The de Sitter solution (39) obtained in imaginary time is invariant with respect to Wick rotation (section 2). Therefore, it is automatically analytically continued into the space of real time where in accordance with (25) it reads $a = a_0 \exp(Ht)$.

## 4. De Sitter state from the classical gravitational waves (CGW)

In this section, we show that Wick rotation applied to the self-consistent set of back reaction equations of CGW and background geometry also produces the de Sitter solution for the empty space as well as in the quantum case considered in sections 2 and 3. The back reaction of CGW of super-long wavelengths on the FLRW background was studied in several papers (see [21] and references therein). In distinction to the quantum case, there is no problem of divergences here but the parameterization problem holds. In these works, the linear parameterization led to the non-conservative EMT and non-self-consistent set of the back reaction equations. Same as in the quantum case, the problem can be solved by exponential parameterization (see [13, section II.D]). The quantum equations (2-6) were obtained by exponential parameterization. We can use them to make the transition to the classical limit. In classical case, they read

$$3H^2 = \kappa \varepsilon_g \qquad (40)$$

$$2\dot{H} + 3H^2 = -\kappa p_g \qquad (41)$$

$$\varepsilon_g = \frac{1}{8\kappa} \sum_{\mathbf{k}\sigma} < \dot{\psi}^+_{\mathbf{k}\sigma} \dot{\psi}_{\mathbf{k}\sigma} + \frac{k^2}{a^2} \psi^+_{\mathbf{k}\sigma} \psi_{\mathbf{k}\sigma} > \qquad (42)$$

$$p_g = \frac{1}{8\kappa} \sum_{\mathbf{k}\sigma} < \dot{\psi}^+_{\mathbf{k}\sigma} \dot{\psi}_{\mathbf{k}\sigma} - \frac{k^2}{3a^2} \psi^+_{\mathbf{k}\sigma} \psi_{\mathbf{k}\sigma} > \qquad (43)$$

$$\phi''_{\bar{k},\sigma} + (k^2 - \frac{a''}{a})\phi_{\bar{k},\sigma} = 0 \qquad \psi_{\mathbf{k}\sigma} = \frac{1}{a}\phi_{\mathbf{k}\sigma} \qquad (44)$$

Equations (40-44) contain functions instead of quantum operators. In imaginary time $\varsigma$, over the de Sitter background (19) the eq. (44) has a solution

$$\psi_{\mathbf{k}\sigma} = \frac{1}{a\sqrt{k}} \left( Q_{\mathbf{k}\sigma} g_k + P_{\mathbf{k}\sigma} h_k \right) \qquad (45)$$

Where $Q_{\mathbf{k}\sigma}$ and $P_{\mathbf{k}\sigma}$ are the integration constants, and mode functions $g_k$ and $h_k$ are (20), i.e. the same as in a quantum case. The requirement of finiteness leads to $P_{\mathbf{k}\sigma} = 0$ in (45). Repeating the calculations of section 2, we get for the flat spectrum of gravitational waves (cf. (23)

$$H_\tau^2 = -\frac{\kappa Q^2 H_\tau^4}{64\pi^2} \tag{46}$$

where $(<|Q_k|^2>_\sigma)^{1/2} \equiv Q = const$. Here $<|Q_k|^2>_\sigma$ is $|Q_k|^2$ averaged over isotropically distributed polarizations. From (46) it follows that

$$H_\tau = -iH \qquad H^2 = 64\pi^2/\kappa Q^2 \tag{47}$$

$$-p_\tau = \varepsilon_\tau = \frac{3Q^2 H_\tau^4}{64\pi^2} \tag{48}$$

$$-p_g = \varepsilon_g = \frac{3Q^2 H^4}{64\pi^2} \tag{49}$$

For the de Sitter state the proper distance to the horizon of events is $H^{-1}$. So, (47) tells us that we deal with the wavelengths of the order of horizon scale. Up to the replacement of constants $Q^2 \to 8\hbar N$, we get the coincidence with the de Sitter solution for quantum gravitons (23-29) which are in accordance with known fact that in quantum terms averaged square of wave amplitude should be proportional to the number of quanta. The same de Sitter solution for CGW can be independently obtained from the BBGKY chain by the same way as it was done in the quantum case [24]. Thus, the Wick rotation provides the self-consistent de Sitter state of the empty Universe formed by CGW as well as quantum gravitons with the same vacuum equation of state.

**5. Gravitons and gravitational waves forming de Sitter solution**

There are no differences between behavior of gravitons and classical gravitational waves described by (44). We're going to continue using the term graviton, bearing in mind that the same behavior applies to the classical gravitational waves. In section 4, the invariance of de Sitter state with respect to Wick rotation provides the analytical continuation of the de Sitter state from imaginary to real time. To complete the analytical continuation of the full self-consistent solution (consisting of the de Sitter background and gravitons forming it), one need to carry out the analytical continuation for graviton mode functions also. The mode functions (15) and (22) are the same for both gravitons and CGW because up to the interchange of operators with functions they are described by the same differential equation (44). Equation (44) can be rewritten in the following form

$$\phi''_{\vec{k},\sigma} + (1 - \frac{a''}{a})\phi_{\vec{k},\sigma} = 0 \tag{50}$$

Here primes mean derivatives over the variable $x = k\eta$. It is similar to the stationary one-dimensional Schrödinger equation. The only differences are that instead of spatial coordinate in the Schrödinger equation, $x$ in (50) is a time coordinate, and $a''/a$ in (50) plays the role of one-dimensional potential.

The solution to (50) over the de Sitter background (13) is (14 and 15) which can be rewritten in the following form

$$\psi_1(x) = C_1(\sin x - x\cos x) + C_2(x\sin x + \cos x) \tag{51}$$

After Wick rotation $x = i\xi$, eq. (50) reads (cf. (18))

$$\phi_{\vec{k},\sigma}'' - (1 + \frac{a''}{a})\phi_{\vec{k},\sigma} = 0 \tag{52}$$

The finite solution to (52) reads (cf. (22))

$$\psi_2(\xi) = C_3(\xi + 1)e^{-\xi} \tag{53}$$

where $C_1$, $C_2$ and $C_3$ are integration constants. To make the analytical continuation (53) to (52), we have to satisfies the following conditions

$$\psi_1(0) = \psi_2(0) \tag{54}$$

$$\psi_1'(0) = \psi_2'(0) \tag{55}$$

Eq. (54) is satisfied by condition $C_2 = C_3$. The (55) is satisfied because

$$\psi_1'(0) = \psi_2'(0) = 0 \tag{56}$$

In terms of quantum tunneling described by the Schrödinger equation, (51) is the sum of incident and reflected waves, (53) is the transition wave and $x = \xi = 0$ is the topologically impenetrable barrier dividing the Lorentzian space of real time and Euclidean space of imaginary time. The latter can be seen from the fact that for $x \ll 1; \xi \ll 1$ equations (50) and (51) are identically the same which means that gravitons of super-horizon wavelengths cease to "feel" any differences between these topologically different spaces and can be belong to both Lorentzian space of real time and Euclidean space of imaginary time. As can be seen from (11) and (17), they are identical for $k = 0$, i.e. in terms of $x$ and $\xi$ the barrier is located at $x = \xi = 0$. In quantum mechanical case, damping of transition wave means the decreasing of probability to find a particle behind the barrier. In distinction to this, in our case the transition gravitational wave is really damped giving its energy to form the de Sitter state behind the barrier which can be seen from (16). Because of invariance of this de Sitter solution with respect to the Wick rotation we can detect it in our Lorentzian space-time. Note that the barrier $x = \xi = 0$ is a boundary between spaces where the signature of the space is changed. The junction conditions on such a boundary are zero derivatives [25], and in accordance with (56) this requirement is satisfied.

## 6. Consistency of de Sitter state formed by gravitons with observational data

*6.1. CMB anisotropy from fluctuations in graviton numbers*

As was shown in [18], the de Sitter solution (26) leads to the fact that fluctuations in the number of gravitons in the Universe reads

$$\frac{<(\Delta N)^2>}{<N>^2} = \frac{1}{8\pi^2} \cdot \frac{H^2}{M_{pl}^2} \qquad (57)$$

Where $M_{pl}$ is the reduced Planck mass $M_{pl} = (\hbar c/8\pi G)^{1/2} = 2.4 \cdot 10^{18} GeV/c^2$. From this it follows that (see (29))

$$\frac{<(\Delta N)^2>}{<N>^2} = \frac{<(\Delta \varepsilon)^2>}{<\varepsilon>^2} \qquad (58)$$

Thus, fluctuations of the number of gravitons produce fluctuations of energy density. They play the same role as scalar perturbations (density fluctuations) which are responsible for the anisotropy of CMB in the models of inflation using scalar fields. In other words, if the graviton condensate is responsible for the inflation then fluctuations of number of gravitons are the cause of the anisotropy of CMB. For the typical energy scale of inflation $H \simeq 10^{15} GeV$ and $\alpha = 1$ one gets from (57) and (58)

$$(\frac{<(\Delta \varepsilon)^2>}{<\varepsilon>^2})^{1/2} \simeq 1.5 \cdot 10^{-5} \qquad (59)$$

As is known, temperature fluctuations $\Delta T/T$ are of the same order of magnitude as the metric and density perturbations which contribute directly to $\Delta T/T$ via the Sachs-Wolfe effect. Thus, the fluctuations of the number of gravitons in the Universe are able to produce CMB anisotropy $\Delta T/T \sim 10^{-5}$ due to fluctuations of gravitational potential which in turn are of the order of fluctuations of energy density. Any mechanism generating the temperature anisotropy inevitably generates the CMB polarization as well, which is an order of magnitude below the temperature fluctuations ([26] and references therein). The B-mode of polarization of CMD is produced by gravitational waves. Thus, both temperature fluctuations and B-mode of polarization of CMB can be of the same origin. Both of them can be produced by primordial gravitons. Thus, the observed CMB anisotropy is consistent with inflation in gravitons of instanton origin.

*6.2. Spectrum of metric fluctuations (inflation and dark energy)*

The de Sitter solution considered above is produced by the flat spectrum of metric fluctuations [18]. The spectrum slightly deviating from the flat spectrum produces the quasi-de Sitter solution. The observed tilt $n_s - 1$ of the power spectrum $k^{n_s - 1}$ deviates slightly from the scale-invariant form corresponding to $n_s = 1$. The observed value is $n_s = 0.96 \pm 0.013$ [27]. This means that in reality we deal with a quasi-de

Sitter expansion. Assuming that $N_k = N_0(k/k_0)^\beta$ where $k_0$ is a pivot scale, we get for $\beta$ in accordance with [18]

$$\beta = 3(1+w) \tag{60}$$

In case of inflation, one starts from the empty space that is gradually filled with the new-born matter. In case of dark energy, one gets the opposite process, one starts from the space filled with matter that is gradually emptied. This leads to the fact that we have to expect to find $w < -1$ for dark energy and $w > -1$ for inflation [18]. In the case of dark energy, the Planck data [6] obtained by a combination of Planck+WP+BAO give, e.g. $w = -1.13^{+0.24}_{-0.25} < -1$. In case of inflation, $n_s - 1 = -\beta$, and for $n_s \approx 0.96$, we get $\beta_{inf} \approx 0.04$ which leads to $w = -0.987 > -1$ [18].

### 6.3. Dark energy

As was already mentioned in Introduction, the dark energy effect was discovered by observations of supernova SNIa by Riess et al. [4] and Perlmutter et al. [5]. Since then a number of hypotheses were advanced to explain this phenomenon (for references to original work, see, e.g. [28-30] and references therein)

#### 6.3.1. Coincidence problem

One of the most intriguing questions in the dark energy effect is the "coincidence problem": Why is the acceleration happening during the contemporary epoch of matter domination? We believe that the answer to this question lies in the combination of two facts which are the existence of a threshold and coincidence of "one-dimensional potentials" for the de Sitter and matter dominated backgrounds [17]. This coincidence distinguishes the matter-dominated epoch from any other. As to the dark energy threshold, in accordance to [17] it should be close to the red shift $z_{threshold} \equiv z_T$

$$1 + z_T \leq (\varepsilon_{de}/\varepsilon_m)^{1/3} = (\Omega_{de}/\Omega_m)^{1/3} \tag{61}$$

The RHS of (61) is presented with generally accepted notation where $\Omega_{de}$ and $\Omega_m$ are ratio of dark energy density and density of matter to the total density of the Universe, respectively, so that $\Omega_{de} + \Omega_m = 1$. Assuming that $\Omega_{de}/\Omega_m \approx 0.69$ [6] one gets from (61) $z_T \leq 0.3$. In the existence of such a threshold lies a possible answer to the question why the birth of dark energy occurs "recently", i.e. after its energy density became comparable with the energy density of matter $\varepsilon_m$.

As to coincidence of "one-dimensional potentials", let us come back to the Schrödinger-like equation (11). For $a = const \cdot \eta^{-\beta}$, the "one-dimensional potential" is $a''/a = \beta(\beta+1)/\eta^2$ (Grishchuk [31]). *The remarkable fact is that "one-dimensional potentials" $a''/a = 2/\eta^2$ are the same in both $\beta = -2$ (matter-dominated background with the equation of state $p = 0$) and $\beta = 1$ (De Sitter background with the equation of state $p = -\varepsilon$) cases.* The same is also true for the imaginary time $\varsigma$. Because "one-dimensional potentials" coincide, the Schrödinger-like equations (11) and (12) for gravitons and ghosts over the matter-dominated background and over the De Sitter background are identical. The same is true of equations (5.1) for CGW. Due to the identity of equations for matter-dominated and De Sitter backgrounds, the boundary conditions for tunneling are naturally satisfied at the barrier. We emphasize once again that this *unique coincidence of "one-dimensional potentials" takes place **only** for the matter dominated epoch and De Sitter State.* In the frame of such reasoning, one can see, e.g. why dark energy was unable to appear during the radiation dominated epoch. The energy eq. (2) in this case reads

$$3H^2 = 8\pi G(\varepsilon_g + \frac{c_r}{a^4} + \frac{c_m}{a^3}) \tag{62}$$

In order that the de Sitter solution could appear it is necessary that the last stage of evolution must be followed by a vacuum stage of empty space. However, the second term of RHS in (62) vanished first, entailing no vacuum but matter-dominated stage of evolution. Therefore, only after the disappearance of the non-relativistic matter (the third term of RHS) can come the vacuum stage, and with it the de Sitter state. Note also that for the radiation-dominated background $a \sim \eta$ there is no barrier for tunneling through because $a''/a = 0$.

6.3.2. The graviton theory of dark energy predicts the correct sign of parameter $1 + w < 0$ which is confirmed by observational data (section 6.2.2).

6.3.3. There are no explicit solutions to the set of equations (2.2-2.6) including the non-relativistic matter to calculate parameter $w$. A typical example of numerical solution to the equations of graviton theory taking into account the input of non-relativistic matter is presented in [17] in Fig.1. It shows that regardless of initial conditions, all solutions come asymptotically to the de Sitter mode that is predicted by graviton theory.

## 7. Conclusion

At the start and by the end of its evolution the Universe is empty (with no matter fields) and dominated by gravitons and classical gravitational waves, respectfully. The back reaction of quantum gravitons as well as classical gravitational waves of super-horizon wavelengths on the background metric leads to the formation of the de Sitter state of the empty Universe. This de Sitter state is the exact solution to the self-consistent equations of one-loop finite quantum gravity as well as the exact solution to the self-consistent equations of back reaction of classical gravitational waves. To get this solution in both quantum and classical cases, one needs to make Wick rotation to imaginary time and back to real time in the process of calculations. The latter means that time was used as a complex variable. The de Sitter state of the Universe at the start and by the end of its evolution looks like a plausible cause of the origin of inflation and dark energy. This theory is consistent with observational data. The cosmological scenario based on the graviton theory was presented in our work [17].

**Acknowledgment**

I am deeply grateful to Daniel Usikov of the University of Maryland for useful discussions. I would like to express my deep appreciation to Walter Sadowski for invaluable help in the preparation of the manuscript. I am grateful to the anonymous referee for comments that helped to improve this paper.